\documentclass{ifacconf}

\usepackage{graphicx}      
\usepackage{natbib}        
\usepackage{tikz}
\usetikzlibrary{patterns}

\newcommand{\nvar}[2]{%
	\newlength{#1}
	\setlength{#1}{#2}
}

\nvar{\dg}{0.3cm}

\nvar{\ddx}{1.5cm}
\nvar{\ddy}{1.5cm}

\usepackage{epstopdf}
\usepackage{xcolor}
\usepackage{amsmath}
\usepackage{amssymb}
\usepackage{pgfplots}
\pgfplotsset{compat=1.7}

\begin{document}
\begin{frontmatter}

\title{Explicit Port-Hamiltonian FEM-Models for Linear Mechanical Systems with Non-Uniform Boundary Conditions}

\thanks[footnoteinfo]{Submitted to 10th Vienna International Conference on Mathematical Modelling}

\author[First]{Tobias Thoma}, 
\author[First]{Paul Kotyczka} 

\address[First]{Technical University of Munich, TUM School of Engineering and Design, Chair of Automatic Control, Garching, Germany (e-mail: tobias.thoma@tum.de)}

\begin{abstract}                
	In this contribution we present how to obtain explicit state space models in port-Hamiltonian form when a mixed finite element method is applied to a linear mechanical system with non-uniform boundary conditions. The key is to express the variational problem based on the principle of virtual power, with both the Dirichlet (velocity) and Neumann (stress) boundary conditions imposed in a weak sense. As a consequence, the formal skew-adjointness of the system operator becomes directly visible after integration by parts, and, after compatible FE discretization, the boundary degrees of freedom of both causalities appear as explicit inputs in the resulting state space model. The rationale behind our formulation is illustrated using a lumped parameter example, and numerical experiments on a one-dimensional rod show the properties of the approach in practice.
\end{abstract}


\end{frontmatter}

\section{Introduction}
\label{sec:introduction}
Port-Hamiltonian (PH) systems provide a framework for modeling, analysis and control of complex dynamical systems where the complexity can result from multi-physical domains and their couplings.
Since several engineering problems are described by partial differential equations the theory of infinite-dimensional PH systems \citep{schaft2002hamiltonian, villegas2007port, jacob2012linear} has become increasingly important in recent years. The progress in modeling different physical domains by means of the PH framework is documented in the review article \cite{Rashad2020}. Also the possibilities and advantages of the PH formulation for structural mechanics have been presented in several articles see e.g. the recent works \cite{Brugnoli2020} and \cite{Warsewa2021}.

For the numerical approximation of distributed parameter PH systems, which is required for simulation and (late lumping) control design, various approaches have been developed in the past 20 years that aim at the preservation of the PH structur, see e.g. the overview in \cite{kotyczka2019numerical}. 
One of the most successful techniques is the partitioned finite element method \citep{farle2013port,Cardoso-Ribeiro2020}. If applied to a system of conservation/balance laws or the variational formulation of an elastic mechanical system, one half of the equations written in a weak form are integrated by parts. The choice of the partially integrated equation determines which boundary condition is satisfied in a weak sense and which type of input variables appears in the resulting finite dimensional control system.

With non-uniform boundary conditions (of both Neumann and Dirichlet type) the approaches presented until now do not immediately lead to an explicit state space model. The first method presented in \cite{Brugnoli2020a} employs Lagrange multipliers and leads to a differential-algebraic finite-dimensional PH system. The second one is based on the virtual decomposition of the domain to interconnect models with different causalities. 
A mixed finite element approach, which leads to the desired explicit state space models with non-uniform causality is introduced in \cite{kotyczka2018weak}. There, however, the choice of original finite element spaces requires (tunable) power-preserving mappings to define the discrete states. 

In this article, we show how to obtain explicit FE models with both inputs and a skew-symmetric interconnection matrix as a visible expression of energy routing in PH systems. We employ the principle of virtual power with weak imposition of boundary conditions, reminiscent of the \emph{Hellinger-Reissner} variational principle \citep{Kaizhou2019}. An advantage of the resulting explicit state space representations is the direct applicability of well-established order reduction techniques for ODE control models.

This article starts with the elastodynamics partial differential equations (PDEs) in Section \ref{sec:2}. In Section \ref{sec:3}, we demonstrate the weak formulation based on the principle of virtual power and the discretization process. In Section \ref{sec:4}, the formulation is illustrated with a one-dimensional rod under Dirichlet-Neumann boundary conditions and its finite-dimensional (approximate) counterpart, a mass-spring chain. Section \ref{sec:5} shows numerical results, and Section \ref{sec:6} gives some short conclusions.

\section{Preliminaries}
\label{sec:2}
This section recalls the first-order (port-)Hamiltonian representation of the linear elastodynamics PDE
\begin{equation}
    \label{eq:elastodyn}
	\rho\ddot{u} - D^TEDu = 0
\end{equation}
on a three dimensional domain $\Omega \subset \mathbb R^3$ in vector notation \citep{Zienkiewicz2005,Brugnoli2020}. The total energy is
\begin{equation}
    \label{eq:Hamiltonian}
	H(u, \dot u) = \frac{1}{2} \int_{\Omega} \dot{u}^T\dot{u}\rho + (Du)^TEDu\;d\Omega 
\end{equation}
with the displacements 
\begin{equation}
	u = \begin{bmatrix} u_1 & u_2 & u_3 \end{bmatrix}^T\in\mathbb{R}^3 
\end{equation}
from an undeformed configuration and the velocities $\dot u \in \mathbb R^3$, the constant mass density ${\rho\in\mathbb{R}}$, the spatial differentiation operator
\begin{equation}
	D = 
	\begin{bmatrix}
	\textstyle\frac{\partial}{\partial x_1} & 0 & 0 \\
	0 & \textstyle\frac{\partial}{\partial x_2} & 0 \\
	0 & 0 & \textstyle\frac{\partial}{\partial x_3} \\
	\textstyle\frac{\partial}{\partial x_2} & \textstyle\frac{\partial}{\partial x_1} & 0 \\
	0 & \textstyle\frac{\partial}{\partial x_3} & \textstyle\frac{\partial}{\partial x_2} \\
	\textstyle\frac{\partial}{\partial x_3} & 0 & \textstyle\frac{\partial}{\partial x_1} 
	\end{bmatrix}
\end{equation}
and the matrix of Hooke's law elastic coefficients
\begin{equation}
	E = 
	\begin{bmatrix}
	\lambda+2G & \lambda & \lambda & 0 & 0 & 0 \\
	\lambda & \lambda+2G & \lambda & 0 & 0 & 0 \\
	\lambda & \lambda & \lambda+2G & 0 & 0 & 0 \\
	0 & 0 & 0 & G & 0 & 0 \\
	0 & 0 & 0 & 0 & G & 0 \\
	0 & 0 & 0 & 0 & 0 & G \\
	\end{bmatrix}\in\mathbb{R}^{6\times 6},
\end{equation}
where $\lambda\in\mathbb{R}$ and $G\in\mathbb{R}$ are known as Lam\'{e} coefficients. Since the strains and stresses are symmetric tensors, they can be represented as a strain vector
\begin{equation}
	\epsilon = \begin{bmatrix}
	\epsilon_{11} & \epsilon_{22} & \epsilon_{33} & \gamma_{12} & \gamma_{23} & \gamma_{13}
	\end{bmatrix}^T\in\mathbb{R}^6
\end{equation}
and a stress vector
\begin{equation}
	\sigma = \begin{bmatrix}
	\sigma_{11} & \sigma_{22} & \sigma_{33} & \sigma_{12} & \sigma_{23} & \sigma_{13}
	\end{bmatrix}^T\in\mathbb{R}^6.
\end{equation}

\emph{Notation.} All field quantities depend on the spatial coordinate $x \in \mathbb R^3$ and the time $t \in \mathbb R$, which we omit for brevity.

\emph{Port-Hamiltonian model.} Defining the vectors of linear momenta $p \in \mathbb R^3$ and strains $\epsilon \in \mathbb R^6$,
\begin{align}
    \label{eq:momenta}
	p &= \rho v \\
	\label{eq:strain}
	\epsilon &= Du 
\end{align}
as energy variables (or states), the Hamiltonian \eqref{eq:Hamiltonian} can be rewritten as $H(p,\epsilon)$. We obtain the fields of co-energy quantities (efforts, co-states) velocity $v \in \mathbb R^3$ and stress $\sigma \in \mathbb R^6$ from $H(p,\epsilon)$ applying the variational derivative, 
\begin{align}
    \label{eq:velocity}
	v = \dot{u} &= \left( \frac{\delta H(p, \epsilon)}{\delta p} \right)^T \\
    \label{eq:stress}
	\sigma = E\epsilon &= \left( \frac{\delta H(p, \epsilon)}{\delta \epsilon} \right)^T.	
\end{align}
Now Eq. \eqref{eq:elastodyn} can be written in a first order co-energy representation
\begin{subequations}
	\label{eq:elasto_ph}
	\begin{align}
	\rho\dot{v} &= D^T\sigma \\[1ex]
	E^{-1}\dot{\sigma} &= Dv,
	\end{align}
\end{subequations}
or summarized,
\begin{equation}
    \label{eq:elasto_ph_matrix}
    \begin{bmatrix}
        \rho I & 0\\ 0 & E^{-1}
    \end{bmatrix}
    \begin{bmatrix}
        \dot v\\ \dot \sigma
    \end{bmatrix}
    =
    \begin{bmatrix}
        0 & D^T\\
        D & 0
    \end{bmatrix}
    \begin{bmatrix}
        v\\ \sigma
    \end{bmatrix},        
\end{equation}
with a s.p.d. matrix on the left, and a formally skew-adjoint differential operator matrix on the right.

\emph{Boundary conditions.} Since this article focuses on non-uniform boundary conditions the boundary ${\partial\Omega = \Sigma_D\cup\Sigma_N}$ is split into two subsets on which the Neumann and Dirichlet boundary condition are applied. The Neumann condition 	
\begin{equation}
    \label{eq:neumann-bc}
	N^T\sigma = \bar{\tau} \quad \text{on}\;\Sigma_N
\end{equation}
with the applied surface traction ${\bar{\tau}\in\mathbb{R}^3}$ and the matrix
\begin{equation}
	N^T = 
	\begin{bmatrix}
	n_1 & 0 & 0 & n_2 & 0 & n_3 \\
	0 & n_2 & 0 & n_1 & n_3 & 0 \\
	0 & 0 & n_3 & 0 & n_2 & n_1 \\
	\end{bmatrix}\in\mathbb{R}^{3\times 6}
\end{equation}
containing the outward normal vector direction components of the surface results form the infinitesimal equilibrium at the surface. The Dirichlet condition
\begin{equation}
    \label{eq:dirichlet-bc}
	v = \bar{\nu} \quad \text{on}\;\Sigma_D
\end{equation}
imposes the desired velocity ${\bar{\nu}\in\mathbb{R}^3}$ on the surface. Differentiating $H(v, \sigma)$ in time, and exploiting the definition of effort variables gives the power balance / conservation law for energy 
\begin{equation}
	\begin{split}
	\dot{H} &= \int_{\partial\Omega}v^TN^T\sigma\;d\partial\Omega \\
	& = \int_{\Sigma_N}v^T\bar{\tau}\;d\Sigma_N + \int_{\Sigma_D}\bar{\nu}^TN^T\sigma\;d\Sigma_D.
	\end{split}
\end{equation}
The surface traction and the velocity field on $\partial \Omega$ are the boundary port variables, their pairing gives the power supplied to the domain $\Omega$. Their causality is opposite on the Neumann and the Dirichlet boundary.

\section{Main results}
\label{sec:3}
In this section we derive the finite dimensional PH system of the elastodynamics equations based on the partitioned finite element method \citep{Cardoso-Ribeiro2020}. By a careful formulation of the problem in weak form we ensure that the structural power balance is immediately visible in the form of the resulting matrices.

\subsection{Weak form}
We start with \eqref{eq:elasto_ph} written in the weak form
\begin{subequations}
	\label{eq:weak}
	\begin{align}
	\begin{split}
	\label{eq:weak_v}
	\delta P_v &= \int_{\Omega} \delta v^T\rho\dot{v} - \delta v^TD^T\sigma\;d\Omega \\
	&\qquad+ \int_{\Sigma_N} \delta v^T(N^T\sigma - \bar{\tau})\;d\Sigma_N = 0 
	\end{split} \\[2ex]
	\begin{split}
	\label{eq:weak_sig}
	\delta P_\sigma &= \int_{\Omega} \delta \sigma^TE^{-1}\dot{\sigma} - \delta \sigma^TDv\;d\Omega \\
	&\qquad+ \int_{\Sigma_D} \delta \sigma^TN(v - \bar{\nu})\;d\Sigma_D = 0
	\end{split}
	\end{align}
\end{subequations}
with the test functions $\delta v$ and $\delta \sigma$, and including the boundary conditions \eqref{eq:neumann-bc}, \eqref{eq:dirichlet-bc}. Equation \eqref{eq:weak_v} represents the residual of the momentum balance equation on $\Omega$ and the Neumann boundary, \eqref{eq:weak_sig} represents the residual of the kinematic equation on $\Omega$ and the Dirichlet boundary. The test functions $\delta v$ and $\delta \sigma$ give a nice interpretation in terms of virtual power, when paired (involving integration) with velocities and stresses. 

\begin{rem} The structure of \eqref{eq:weak} is reminiscent of the \textit{Hellinger-Reissner} principle, where Dirichlet boundary conditions can also be imposed weakly \citep{Kaizhou2019}.
\end{rem}

\subsection{Discretization}
\begin{thm}
    The mixed Galerkin discretization of \eqref{eq:elasto_ph_matrix} with Neumann and Dirichlet boundary conditions \eqref{eq:neumann-bc} and \eqref{eq:dirichlet-bc} based on the weak formulation \eqref{eq:weak} and using trial and test functions from the same bases,
    \begin{align}
        \label{eq:test-trial-v}
    	v(x,t) &= \phi(x)^T\hat{v}(t), & 
    	 \delta v(x) &= \phi(x)^T\delta\hat{v},
    	\\[1ex]
    	\label{eq:test-trial-sigma}
    	\sigma(x,t) &= \psi(x)^T\hat{\sigma}(t), &
    	\delta \sigma(x) &= \psi(x)^T\delta\hat{\sigma}, 
    	\\[1ex]
    	\bar{\tau}(x,t) &= \xi(x)^T\hat{\tau}(t), &
    	 \bar{\nu}(x,t) &= \zeta(x)^T\hat{\nu}(t),
    \end{align}    
    leads to the explicit PH state space model in co-energy form\footnote{Here, $\hat u$ denotes the input vector, and not the displacement degrees of freedom. The representation is explicit, because $M$ is invertible.}
    \begin{align}
	\label{eq:disc_sys}
	\underbrace{
	\begin{bmatrix}
	M_v & 0 \\
	0 & M_\sigma
	\end{bmatrix}}_{M}
	\underbrace{
	\begin{bmatrix}
	\dot{\hat{v}} \\
	\dot{\hat{\sigma}}
	\end{bmatrix}}_{\dot{\hat{e}}} &=
	\underbrace{
	\begin{bmatrix}
	0 & -K \\
	K^T & 0
	\end{bmatrix}}_{J}
	\underbrace{
	\begin{bmatrix}
	\hat{v} \\
	\hat{\sigma}
	\end{bmatrix}}_{\hat{e}} +
	\underbrace{
	\begin{bmatrix}
	G_v & 0 \\
	0 & G_\sigma
	\end{bmatrix}}_{G}
	\underbrace{
	\begin{bmatrix}
	\hat{\tau} \\
	\hat{\nu}
	\end{bmatrix}}_{\hat{u}}\\
	\label{eq:disc-output}
    \hat{y} &=
    \underbrace{
    \begin{bmatrix}
	G_v^T & 0 \\
	0 & G_\sigma^T
	\end{bmatrix}}_{G^T}
	\underbrace{
	\begin{bmatrix}
	\hat{v} \\
	\hat{\sigma}
	\end{bmatrix}}_{\hat{e}},
\end{align}
where $\hat e = \nabla \hat H$ with the approximate Hamiltonian
\begin{equation}
    \label{eq:Hamiltonian-discrete}
    \hat H = \frac{1}{2} \hat e^T M \hat e.
\end{equation}
The sub-matrices are given by 
\begin{align}
    \label{eq:M_v}
	M_v &= \int_{\Omega} \phi \rho \phi^T\;d\Omega \\
	M_\sigma &= \int_{\Omega}\psi E^{-1}\psi^T\;d\Omega \\
	K &= \int_{\Omega}(D\phi^T)^T\psi^T\;d\Omega - \int_{\Sigma_D}\phi N^T\psi^T\;d\Sigma_D \\
	G_v &= \int_{\Sigma_N}\phi\xi^T\;d\Sigma_N \\
	\label{eq:G_sigma}
	G_\sigma &= \int_{\Sigma_D}\psi N\zeta^T\;d\Sigma_D.
\end{align}
\end{thm}

\begin{pf}
    After integration by parts of \eqref{eq:weak_v}, the weak form can be written as
	\begin{subequations}
		\begin{align}
		\nonumber
		\delta P_v &= \int_{\Omega} \delta v^T\rho\dot{v} + (D\delta v)^T\sigma\;d\Omega  - \int_{\partial\Omega} \delta v^TN^T\sigma\;d\partial\Omega \\
		\label{eq:weak_1}
		&\qquad + \int_{\Sigma_N} \delta v^T(N^T\sigma - \bar{\tau})\;d\Sigma_N = 0,
         \\[2ex]
        \nonumber
		\delta P_\sigma &= \int_{\Omega} \delta \sigma^TE^{-1}\dot{\sigma} - \delta \sigma^TDv\;d\Omega \\
		\label{eq:weak_2}
		&\qquad+ \int_{\Sigma_D} \delta \sigma^TN(v - \bar{\nu})\;d\Sigma_D = 0.
        \end{align}
	\end{subequations}	
	Splitting the second integral in \eqref{eq:weak_1} into the parts on $\Sigma_N$ and $\Sigma_D$, we get
	\begin{subequations}
		\label{eq:weak_ibp}
		\begin{align}
		\label{eq:weak_ibp_v}
% 		\begin{split}
		\delta P_v &= \underbrace{\int_{\Omega} \delta v^T\rho\dot{v}\;d\Omega + \int_\Omega (D\delta v)^T\sigma\;d\Omega}_{-\delta P_{\text{int}}} \\ \nonumber
		&\qquad \underbrace{- \int_{\Sigma_D} \delta v^TN^T\sigma\;d\Sigma_D 
			- \int_{\Sigma_N} \delta v^T\bar{\tau}\;d\Sigma_N}_{-\delta P_{\text{ext}}} = 0,\\
% 		\end{split} \\
		\label{eq:weak_ibp_sig}
% 		\begin{split}
		\delta P_\sigma &= \int_{\Omega} \delta \sigma^TE^{-1}\dot{\sigma}\;d\Omega - \int_\Omega \delta \sigma^TDv\;d\Omega \\ \nonumber
		&\qquad+ \int_{\Sigma_D} \delta \sigma^TN v\;d\Sigma_D - \int_{\Sigma_D} \delta \sigma^TN \bar{\nu}\;d\Sigma_D = 0.
% 		\end{split}
		\end{align}
	\end{subequations}
The second and third terms already suggest the appearance of a (formally) skew-adjoint operator, which will turn to a skew-symmetric interconnection matrix in the discretized model.

\eqref{eq:weak_ibp_v} represents the well-known balance between the negative internal virtual $\delta P_{\text{int}}$ and external power $\delta P_{\text{ext}}$. $\delta P_\text{int}$ includes the virtual power due to inertia and virtual stress power. $\delta P_\text{ext}$ describes the virtual power due to external forces. \eqref{eq:weak_ibp_sig} follows by the principle of virtual forces and fulfills the kinematic equation and the Dirichlet condition. Because the vector $\bar{\nu}$ due to the Dirichlet boundary conditions is applied in a weak sense, the space of virtual velocities $\delta v$ will not be restricted due to the constraints resulting from the Dirichlet boundary conditions. Therefore, the term ${\int_{\Sigma_D} \delta v^TN^T\sigma\;d\Sigma_D \neq 0}$ representing reaction forces -- being compatible with the Dirichlet constraints -- does not vanish compared to the case of strong Dirichlet conditions, where ${\int_{\Sigma_D} \delta v^TN^T\sigma\;d\Sigma_D = 0}$ due to the principle of d'Alembert. Section \ref{sec:4} will illustrate this idea using a simple example.

Inserting the test and trial functions according to \eqref{eq:test-trial-v}, \eqref{eq:weak_ibp} turns to
\begin{subequations}
		\begin{align}
		\begin{split}
		\delta P_v &= \int_{\Omega} \delta \hat{v}^T\phi\rho\phi^T\dot{\hat{v}} +
		\delta\hat{v}^T(D \phi^T)^T\psi^T\hat{\sigma}\;d\Omega \\
		&\qquad - \int_{\Sigma_D} \delta \hat{v}^T\phi N^T\psi^T\hat{\sigma}\;d\Sigma_D \\
		&\qquad	- \int_{\Sigma_N} \delta \hat{v}^T\phi\xi^T\hat{\tau}\;d\Sigma_N = 0,
		\end{split} \\[2ex]
		\begin{split}
		\delta P_\sigma &= \int_{\Omega} \delta \hat{\sigma}^T\psi E^{-1}\psi^T\dot{\hat{\sigma}} 
		- \delta \hat{\sigma}^T \psi (D\phi^T) \hat{v}\;d\Omega \\
		&\qquad+ \int_{\Sigma_D} \delta \hat{\sigma}^T\psi N\phi^T \hat{v}\;d\Sigma_D \\
		&\qquad- \int_{\Sigma_D} \delta \hat{\sigma}^T\psi N \zeta^T\hat{\nu}\;d\Sigma_D = 0.
		\end{split}
		\end{align}
	\end{subequations}
Since all quantities not depending on the spatial coordinates $x$ can be taken out of the integral, and the equations must hold for all ${\delta \hat{v}}$ and ${\delta \hat{\sigma}}$, we obtain the resulting set of ODEs \eqref{eq:disc_sys} with matrices \eqref{eq:M_v}--\eqref{eq:G_sigma}.

Inserting the approximation of the co-energy variables in ${H(u,\dot u)}$ according to \eqref{eq:Hamiltonian} with $E D u = \sigma$ and $\dot u = v$, yields
\begin{equation}
    \hat{H} = \frac{1}{2}\int_{\Omega} \hat{v}^T\phi \rho \phi^T\hat{v} + \hat{\sigma}^T\psi E^{-1}\psi^T\hat{\sigma}\;d\Omega,
\end{equation}
or, written in compact form, \eqref{eq:Hamiltonian-discrete}.

From the time derivative of the discretized energy,
\begin{equation}
		\dot{\hat H} = \hat{v}^TG_v\hat{\tau} + \hat{\sigma}^TG_\sigma \hat{\nu},
	\end{equation}
we obtain the natural definition of the discrete power-conjugated outputs \eqref{eq:disc-output}.
\end{pf}

\begin{rem}
	Integration by parts of \eqref{eq:weak_sig} is also an option to get a discrete PH system. Nevertheless, the spatial derivative operator $D$ restricts the choice of the basis functions for $\sigma$ in this case.
\end{rem}

\begin{rem}
    A mixed formulation \eqref{eq:weak} can lead to higher accuracy compared to single field principles \citep{Zienkiewicz2005}.
\end{rem}

\section{A one-dimensional example}
\label{sec:4}

To illustrate that the approach of weak imposition of both types of boundary conditions has an intuitive interpretation also in finite dimension, we consider a one dimensional rod and the depicted mass-damper chain (with $N=2$ springs for simplicity) as its finite-dimensional counterpart.

\subsection{One-dimensional rod}
First we consider a rod of length $L$
\begin{subequations}
	\begin{align}
	\label{eq:04:0010a}
	\rho\dot{v} &= \frac{\partial \sigma}{\partial x} \\[1ex]
	\label{eq:04:0010b}
	\frac{1}{EA}\dot{\sigma} &= \frac{\partial v}{\partial x}
	\end{align}
\end{subequations}
with the energy variables momentum density $p = \rho v$ and strain $\varepsilon = \sigma/(EA)$ and the co-energy variables velocity $v$ and normal force $\sigma$. The boundary conditions are
\begin{subequations}
\begin{align}
    \label{eq:04:0020a}
    v(x=0,t) &= \bar{\nu}\\[1ex]
    \label{eq:04:0020b}    
    \sigma(x=L,t) &= \bar{\tau}.
\end{align}
\end{subequations}
Imposing them in weak form as described in {Section 3.1} gives us for all variations $\delta v(x)$ and $\delta \sigma(x)$
\begin{align}
    \label{eq:04:0030}
	\delta P_v &= \int_{0}^L \delta v \rho \dot{v} + \frac{\partial \delta v}{\partial x} \sigma \;dx + 
	\delta v(0) \sigma(0) - \delta v(L) \bar{\tau} = 0,\\
	\label{eq:04:0040}
	\delta P_\sigma &= \int_{0}^L \delta \sigma \frac{\dot{\sigma}}{EA} - \delta \sigma \frac{\partial v}{\partial x} \;dx + \delta F(0)\cdot(\bar{\nu} - v(0)) = 0,
\end{align}
which has a nice connection to the following finite-dimensional system.

\subsection{Spring-mass chain}

\begin{figure}[h]
    \centering
	%%% two mass-sprig system start %%%
\def\mh{0.5}
\def\doublemass{%
	% mass one
	\draw[very thick] (0,0) circle (0.5*\mh);
	% degree of freedom one 
	\begin{scope}[red]
		\draw [dashed] (0,0) -- (0pt,3*\mh);
		\draw [->] (-1.5*\mh,2*\mh) -- (0,2*\mh);
		\draw (-1.5*\mh,2*\mh) -- ++ (0,3pt);
		\draw (-1.5*\mh,2*\mh) -- ++ (0,-3pt);
		\node at (-1.2*\mh,2.6*\mh) {$v_0 = \bar{\nu}$};
	\end{scope}
	\begin{scope}[blue]
		\draw [dashed] (0,0) -- (0pt,-3*\mh);
		\draw [<->] (0*\mh,-2*\mh) -- (5*\mh,-2*\mh);
		\node at (2.5*\mh,-1.6*\mh) {$\Delta q_1 = q_1 - q_0$};
	\end{scope}
	\node at (1.5*\mh,-0.7*\mh) {$m_0=0$};
	% mass two
	\begin{scope}[shift=(0:5*\mh), rotate=0]
		\draw[very thick, fill] (0,0) circle (0.5*\mh);
		% degree of freedom two
		\begin{scope}[red]
			\draw [dashed] (0,0) -- (0pt,3*\mh);
			\draw [->] (-1.5*\mh,2*\mh) -- (0,2*\mh);
			\draw (-1.5*\mh,2*\mh) -- ++ (0,3pt);
			\draw (-1.5*\mh,2*\mh) -- ++ (0,-3pt);
			\node at (-0.8*\mh,2.4*\mh) {$v_1$};
		\end{scope}
		\begin{scope}[blue]
			\draw [dashed] (0,0) -- (0pt,-3*\mh);
			\draw [<->] (0*\mh,-2*\mh) -- (5*\mh,-2*\mh);
			\node at (2.5*\mh,-1.6*\mh) {$\Delta q_2 = q_2 - q_1$};
			\draw [dashed] (5*\mh,0) -- (5*\mh,-3*\mh);
		\end{scope}
		% spring
		\draw [spring] (-4*\mh,0) -- (-\mh,0);
		\node at (-5*\mh*\mh,0.5*\mh) {$c_1$};
		\node at (2.5*\mh,0.5*\mh) {$c_2$};
		\node at (\mh,-0.7*\mh) {$m_1$};
	\end{scope}
	% mass three
	\begin{scope}[shift=(0:10*\mh), rotate=0]
		\draw[very thick, fill] (0,0) circle (0.5*\mh);
		% degree of freedom two
		\begin{scope}[red]
			\draw [dashed] (0,0) -- (0pt,3*\mh);
			\draw [->] (-1.5*\mh,2*\mh) -- (0,2*\mh);
			\draw (-1.5*\mh,2*\mh) -- ++ (0,3pt);
			\draw (-1.5*\mh,2*\mh) -- ++ (0,-3pt);
			\node at (-0.8*\mh,2.4*\mh) {$v_2$};
		\end{scope}
		% spring
		\draw [spring] (-4*\mh,0) -- (-\mh,0);
		\node at (\mh,-0.7*\mh) {$m_2$};
	\end{scope}
	}
%%% two mass-sprig system end %%%

\begin{tikzpicture}[scale = 1.0]
\tikzstyle{spring}=[thick,decorate,decoration={zigzag,pre length=0.3cm,post
	length=0.3cm,segment length=6}]
\tikzstyle{damper}=[thick,decoration={markings,  
	mark connection node=dmp,
	mark=at position 0.5 with 
	{
		\node (dmp) [thick,inner sep=0pt,transform shape,rotate=-90,minimum
		width=10pt,minimum height=3pt,draw=none] {};
		\draw [thick] ($(dmp.north east)+(2pt,0)$) -- (dmp.south east) -- (dmp.south
		west) -- ($(dmp.north west)+(2pt,0)$);
		\draw [thick] ($(dmp.north)+(0,-3pt)$) -- ($(dmp.north)+(0,3pt)$);
	}
}, decorate]
% uncontrolled system
\doublemass

\begin{scope}[blue]
\draw [very thick, ->] (0.5*\mh,0) -- (1.5*\mh,0);
\node at (\mh,0.5*\mh) {$F_1$};
\draw [very thick, ->] (-0.5*\mh,0) -- (-1.5*\mh,0);
\node at (-\mh,0.5*\mh) {$F_0$};
\end{scope}

\begin{scope}[shift=(0:5*\mh), rotate=0]
\begin{scope}[blue]
\draw [very thick, ->] (0.5*\mh,0) -- (1.5*\mh,0);
\node at (\mh,0.5*\mh) {$F_2$};
\draw [very thick, ->] (-0.5*\mh,0) -- (-1.5*\mh,0);
\node at (-\mh,0.5*\mh) {$F_1$};
\end{scope}
\end{scope}

\begin{scope}[shift=(0:10*\mh), rotate=0]
\begin{scope}[blue]
\draw [very thick, ->] (0.5*\mh,0) -- (1.5*\mh,0);
\node at (2.8*\mh,0) {$F_3=\bar{\tau}$};
\draw [very thick, ->] (-0.5*\mh,0) -- (-1.5*\mh,0);
\node at (-\mh,0.5*\mh) {$F_2$};
\end{scope}
\end{scope}

\end{tikzpicture}
	\caption{Spring-mass chain with $N=2$}
	\label{fig:mass-spring-chain}
\end{figure}
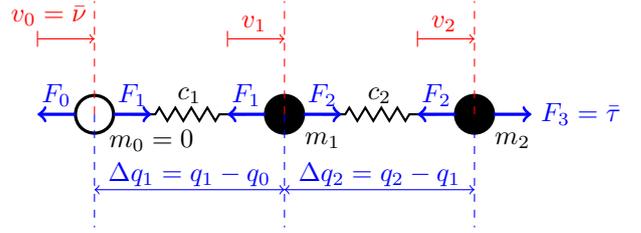

For simplicity we consider $N = 2$ springs and $N+1$ (mass) points at their terminals, see Fig. \ref{fig:mass-spring-chain}. The argumentation holds for arbitrary $N$.

\emph{Dirichlet and Neumann BCs.} If we impose a Dirichlet BC at a terminal point (here the left one, index $0$), its mass is not considered in the lumped kinetic energy. The reason is that the velocity of the (mass) point is not derived from its kinetic energy, but directly imposed as a BC. For Neumann BC (here the right terminal), the input force specifies the value of the external force $F_{N+1} = F_3$, which contributes to the acceleration of the corresponding mass.

\emph{States and co-states.} With this convention we first set up the canonical states, the energies and the co-states for masses and springs as storage elements for kinetic and potential energy. With $p_1$, $p_2$ the momenta of the masses, and $\Delta q_1 = q_1 - q_0$, $\Delta q_2 = q_2 - q_1$ the elongations of the springs from their undeformed state, the total energy is
\begin{equation}
	\label{eq:exa:0010}
	H(p, \Delta q) = \sum_{i=1}^2 \frac{1}{2m_i}p_i^2 + \sum_{i=1}^2 \frac{1}{2}c_i \Delta q_i^2,
\end{equation}
with $m_i$ and $c_i$, $i=1,2$, masses and stiffnesses. The velocities $v_1$, $v_2$ and the restoring spring forces $F_1$, $F_2$, as shown in Fig. \ref{fig:mass-spring-chain}, are co-state or effort variables that are derived from the Hamiltonian:
\begin{equation}
	\label{eq:exa:0020}
	v_i = \frac{\partial H}{\partial p_i}, \quad 
	F_i = \frac{\partial H}{\partial \Delta q_i}, \qquad i=1,2.
\end{equation}

\emph{Virtual power based on velocity variations.} We express two formulations of the principle of virtual power. The first is the virtual power balance based on virtual velocities $\delta v_0$, $\delta v_1$, $\delta v_2$ (we allow for $\delta v_0 \neq 0$, as $v_0$ is imposed only \emph{weakly}):
\begin{equation}
	\label{eq:exa:0030}
	\delta P_v = - \delta P^m + \delta P^v_{ic} = 0,
\end{equation}
where 
\begin{equation}
	\label{eq:exa:0040}
	\begin{split}
	\delta P^m &= \dot p_1 \delta v_1 + \dot p_2 \delta v_2\\
		&= - f_1^m \delta e_1^m - f_1^m \delta e_2^m
	\end{split}
\end{equation}
expresses the variation of power supplied to the kinetic energy storage elements (masses), and $f_i^m = - \dot p_i$, $i=1,2$ are the canonical flows using a generator sign convention. On the contrary,
\begin{equation}
	\label{eq:exa:0050}
	\delta P_{ic}^v = (F_1 - F_0) \delta v_0 + (F_2 - F_1) \delta v_1 + (F_3 - F_2) \delta v_2
\end{equation}
is the variation of the power injected to the remaining system and its environment at the locations of the masses. Specifying the right boundary force in a weak sense,
\begin{equation}
	\label{eq:exa:0060}
	(\bar{\tau} - F_3) \delta v_2 = 0 \qquad \forall\; \delta v_2,
\end{equation}
and adding this expression to \eqref{eq:exa:0030}, we obtain after sorting terms,
\begin{multline}
	\label{eq:exa:0070}
	\delta P_v = (F_0 - F_1) \delta v_0\\ + (\dot p_1 + F_1 - F_2) \delta v_1 + (\dot p_2 + F_2 - \bar{\tau}) \delta v_2 = 0.
\end{multline}
Requiring the expression to vanish for all velocity variations, we recover the ODEs for the momenta of the masses and the expression for the reaction force at the left boundary. Note on the other hand that we can re-sort the terms on the right as
\begin{equation}
    \label{eq:exa:0075}
    \delta P_v = \sum_{i=1}^2 ( \delta v_i \dot p_i + \Delta \delta v_i F_i ) + \delta v_0 F_0 - \delta v_2 \bar{\tau} = 0
\end{equation}
with $\Delta \delta v_i = \delta v_{i+1}-\delta v_i$, which is the discrete version of \eqref{eq:04:0030}.

\emph{Virtual power based on force variations.} We now consider the variations $\delta F_1$, $\delta F_2$ of the restoring  forces (which originate in variations $\delta \Delta q_1$, $\delta \Delta q_2$) and the resulting virtual power balance
\begin{equation}
	\label{eq:exa:0080}	
	\delta P_F = - \delta P^c + \delta P_{ic}^F = 0.
\end{equation}
The term
\begin{equation}
	\label{eq:exa:0090}	
	\begin{split}
	\delta P^c &= \Delta \dot q_1 \delta F_1 + \Delta \dot q_2 \delta F_2\\
	&= -f_1^c \delta e_1^c - f_2^c \delta e_2^c
	\end{split}
\end{equation}
represents the power variation supplied to the springs, while 
\begin{equation}
	\label{eq:exa:0100}
	\delta P_{ic}^F = - v_0 \delta F_1 + v_1( \delta F_1 - \delta F_2) + v_2 \delta F_2
\end{equation}
is the power variation flowing to the rest of the system at the nodes. Adding the velocity boundary condition in weak form
\begin{equation}
	\label{eq:exa:0110}	
	(v_0 - \bar{\nu}) \delta F_1 = 0 \qquad \forall\; \delta F_1,
\end{equation}
to \eqref{eq:exa:0080}, we obtain (after a change of sign)
\begin{equation}
    \label{eq:exa:0130}	 
    \delta P_F = \sum_{i=1}^2 \delta F_i \Delta \dot{q}_i - \delta F_i (v_{i}-v_{i-1}) + \delta F_i( \bar{\nu} - v_0) = 0, 
\end{equation}
which is the discrete version of \eqref{eq:04:0040}. Again, we re-sort the terms and obtain
\begin{equation}
    \label{eq:exa:0120}	
	\delta P_F = (\Delta \dot q_1 - v_1 + \bar{\nu}) \delta F_1 + (\Delta \dot q_2 + v_1 - v_2) \delta F_2 = 0,
\end{equation}
from which, for arbitrary variations $\delta F_1$, $\delta F_2$ the ODEs for the elongations $\Delta q_1$, $\Delta q_2$ follow. With $\dot q_0 = v_0 = \bar{\nu}$ and the initial values of $q_0$, $q_1$ and $q_2$, the motion of the system is completely determined. 

\section{Numerical results}
\label{sec:5}
The performance of our approach is now demonstrated with a simple FEniCS simulation \citep{LoggMardalEtAl2012}.
Therefore, the one dimensional rod of Section \ref{sec:4}.1
with the initial conditions
\begin{align}
v(x,t=0) &= 0\;\text{m/s} \\[1ex]
\sigma(x,t=0) &= 0\;\text{N}
\end{align}
is our benchmark system. The rod is first discretized with 100 and then with 200 elements, second order Lagrange polynomials for $\phi$ and first order discontinuous basis functions for $\psi$ leading in case of 100 elements to ${\hat{v}\in\mathbb{R}^{201}}$ and ${\hat{\sigma}\in\mathbb{R}^{200}}$. This common choice results from a classic finite element approach for mechanical systems, where the displacements are usually discretized with Lagrange polynomials of order $\delta$ leading to discontinuous stresses of order $\delta-1$ (see. \eqref{eq:strain} and \eqref{eq:stress}).

The benchmark system is simulated for ${T=10\;\text{ms}}$ using the implicit midpoint rule and sampling time ${\Delta T = 1\cdot 10^{-3}\;\text{ms}}$. Further simulation parameters are listed in Table \ref{tab:sim}.

\begin{table}[htbp]
	\centering
	\caption{Simulation parameters} 
	\begin{tabular}{ c | c }
		Symbol & Value  \\
		\hline
		$L$ & $1\;\text{m}$ \\ 
		$\rho$ & $0.7850\;\text{kg/m}$ \\ 
		$E$ & $200\cdot 10^3\:\text{N/mm}^2$ \\
		$A$ & $100\:\text{mm}^2$ \\
		$\bar{\nu}$ & $0\;\text{m/s}$ \\
		$\bar{\tau}$ & $\forall t\leq0.5\;\text{ms}\quad\tau=1000\;\text{N}\quad\text{else}\quad \tau=0\;\text{N}$\\
		\hline
	\end{tabular}
	\label{tab:sim}
\end{table}

\begin{figure}[htbp]
	\centering
	\includegraphics[width =0.45\textwidth]{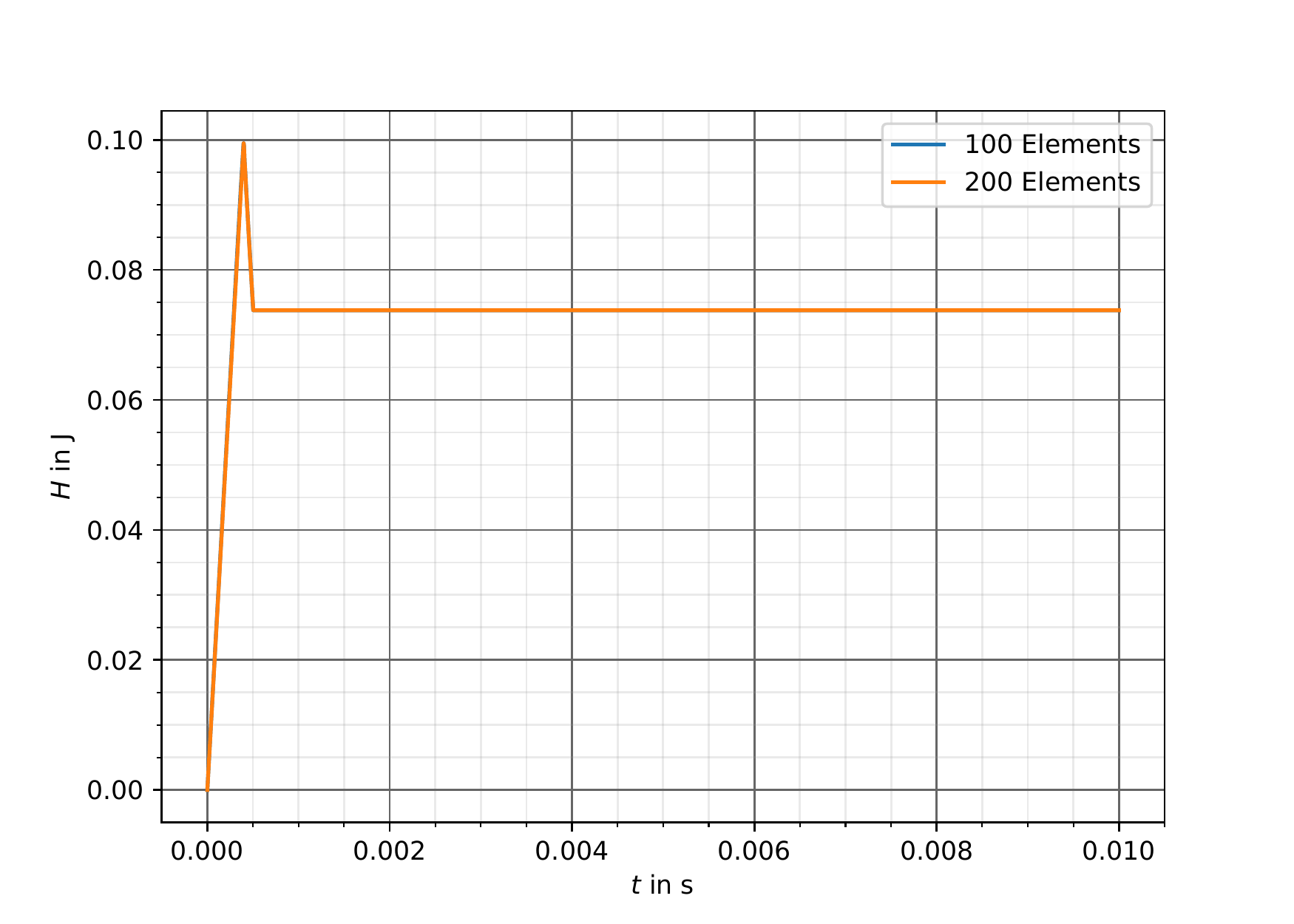}
	\caption{Hamiltonian $\bar H$ of the rod}
	\label{fig:H}
\end{figure}

\begin{figure}[htbp]
	\centering
	\includegraphics[width =0.45\textwidth]{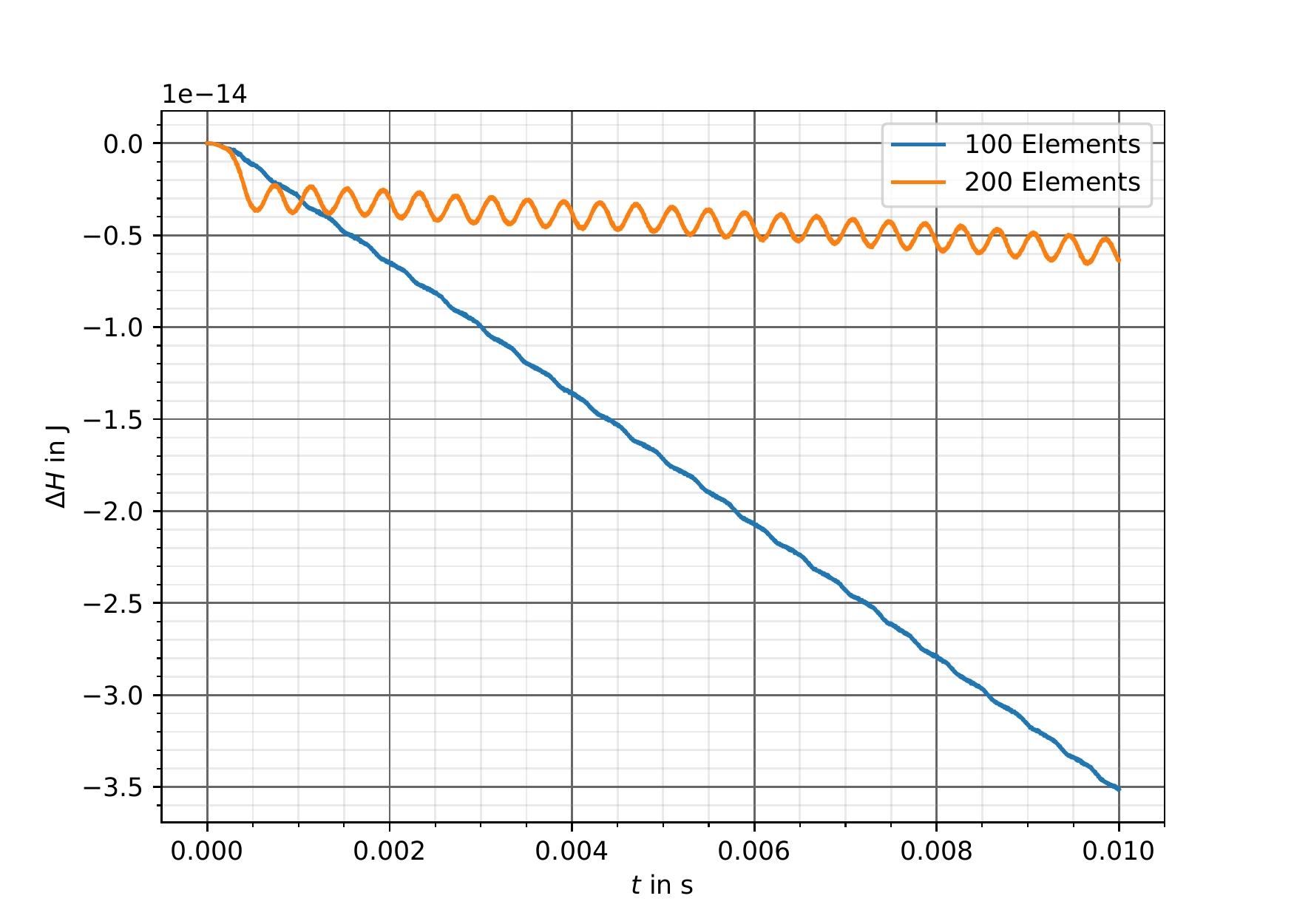}
	\caption{Energy residual $\Delta H$ according to \eqref{eq:H-residual}}
	\label{fig:res}
\end{figure}

\begin{figure}[htbp]
	\centering
	\includegraphics[width =0.45\textwidth]{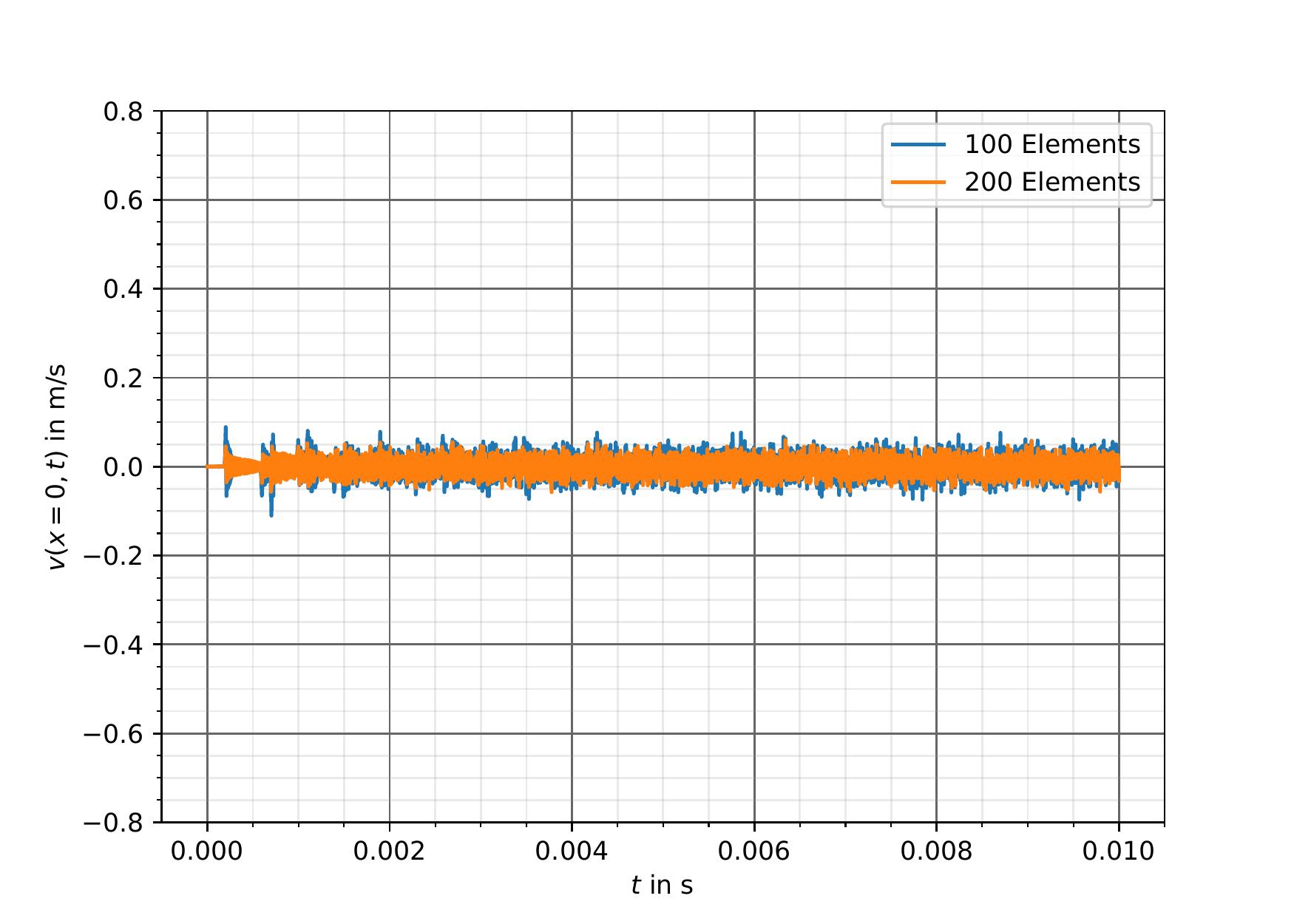}
	\caption{Dirichlet boundary condition. $v(x=0,t)$ meets $\bar{\nu}=0$ in a weak sense.}
	\label{fig:v}
\end{figure}

\begin{figure}[htbp]
	\centering
	\includegraphics[width =0.45\textwidth]{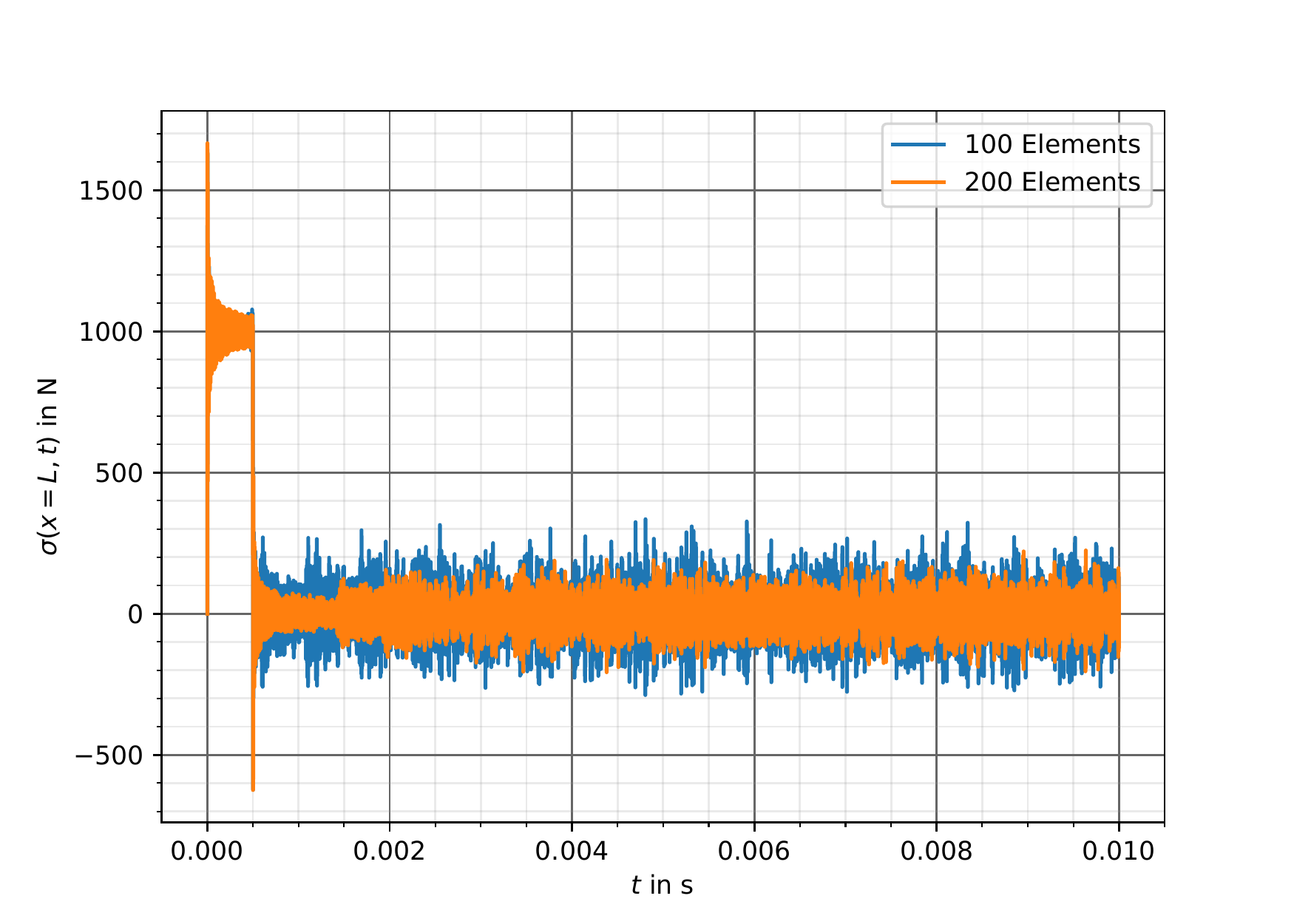}
	\caption{Neumann boundary condition. $\sigma(x=L,t)$ meets $\bar{\tau}$ according to Table \ref{tab:sim} in a weak sense.}
	\label{fig:tau}
\end{figure}

\begin{figure}[htbp]
	\centering
	\includegraphics[width =0.45\textwidth]{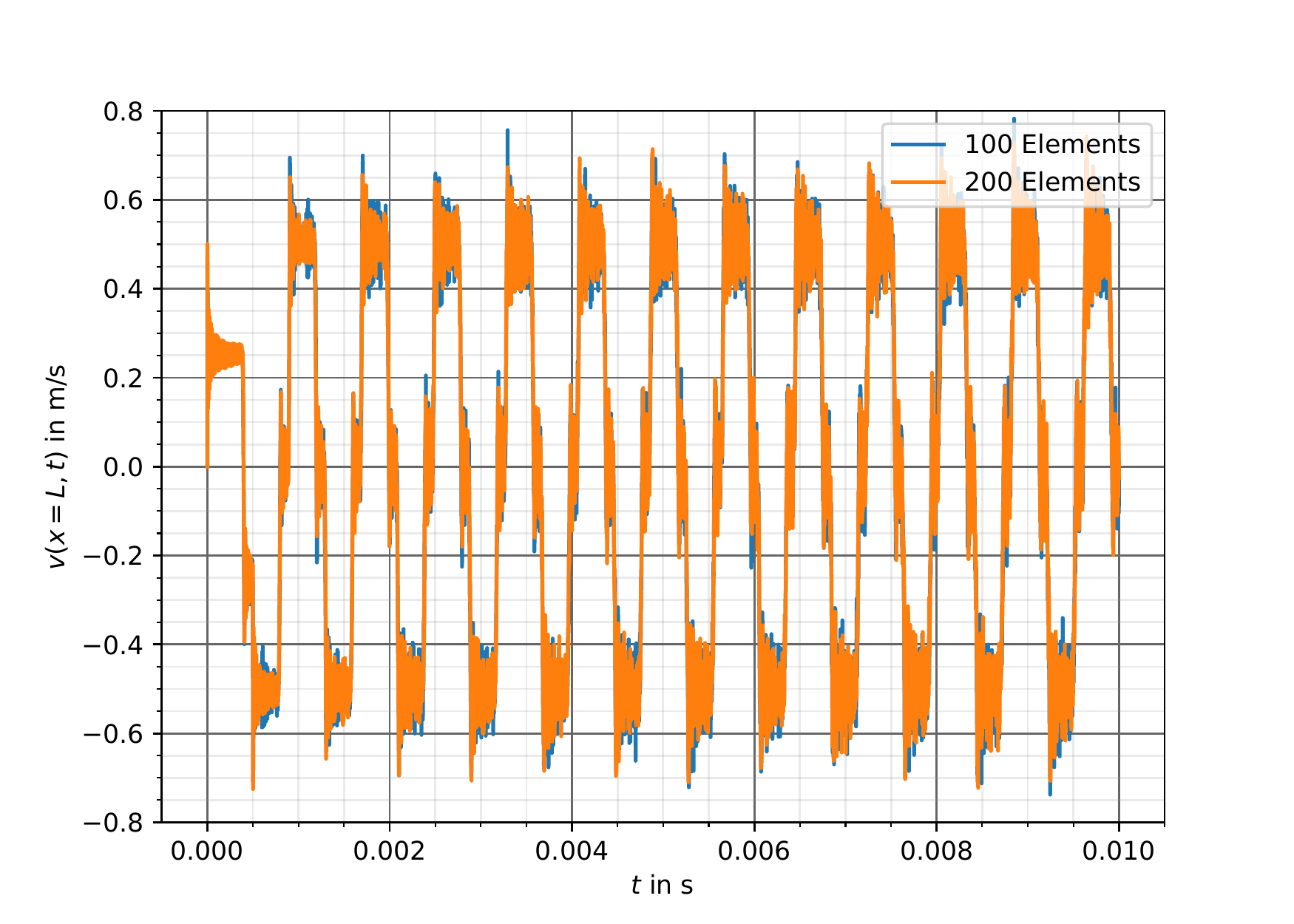}
	\caption{Velocity $v(x=L,t)$}
	\label{fig:v_tip}
\end{figure}

Fig. \ref{fig:H} shows the Hamiltonian of the rod, which remains as expected constant for $t>0.5\;\text{ms}$ up to machine precision. The energy introduced into the system results from the Neumann boundary condition. The energy residual (in the approximation spaces)
\begin{equation}
    \label{eq:H-residual}
\Delta H = H - \int_{0}^{t}\bar{\nu}\cdot\sigma(x=0,t)\;dt - \int_{0}^{t}v(x=L,t)\cdot\bar{\tau}\;dt
\end{equation}
is shown in Fig. \ref{fig:res}. Figures \ref{fig:v} and \ref{fig:tau} illustrate compliance with the boundary conditions in a weak sense. For the sake of completeness, Fig. \ref{fig:v_tip} shows the velocities at $x=L$.

Another point worth mentioning is the resulting eigenvalues of the discrete system compared to the exact ones.
The first six scaled eigenvalues ${\lambda=\frac{\rho}{EA}\omega^2}$, where $\omega$ represents the eigenfrequencies of the conjugate complex eigenvalue pairs, are clarified in Table \ref{tab:eig}. The discretized system \eqref{eq:disc_sys} has an additional eigenvalue in 0 due to the odd dimension of the skew-symmetric interconnection matrix $J$. This eigenvalue/-vector establishes an invariant out of a combination of some node velocities $\hat{v}$.

\begin{table}[htbp]
	\centering
	\caption{Scaled eigenvalues $\lambda = \frac{\rho}{EA}\omega^2$} 
	\begin{tabular}{ c | c | c}
		Number & computed (100 elements) & exact \\		
		\hline
		1 & 0 & - \\
		2 & 2.4674 & 2.4674 \\
		3 & 22.2067 & 22.2066 \\
		4 & 61.6854 & 61.6850 \\
		5 & 120.9042 & 120.9027 \\
		6 & 199.8637 & 199.8595 \\
		\hline
	\end{tabular}
	\label{tab:eig}
\end{table}

\section{Conclusions}
\label{sec:6}
We presented a systematic approach for structure-preserving discretization of infinite dimensional mechanical systems with non-uniform boundary conditions. Our approach is motivated by the principle of virtual power and can be easily extended to other systems like systems of conservation/balance laws. The explicit nature of the resulting finite-dimensional models is a valuable property for their further control-oriented treatment, e.g., structure-preserving order reduction.

\begin{ack}
	The authors thank Boris Lohmann, Tim Moser and Christopher Lerch for fruitful discussions.
\end{ack}

\bibliography{2021_ThomKoty_ActaMechanica}            

\end{document}